\documentclass[a4paper, twocolumn]{article}
\usepackage[utf8]{inputenc}
\usepackage{cite}
\usepackage{graphicx}
\usepackage{amssymb}
\usepackage{amsfonts}
\usepackage{amsmath}
\usepackage{bm}
\usepackage{longtable}
\usepackage{epsfig}
\usepackage{multirow}
\usepackage{authblk}%
\usepackage[paperwidth=19.5cm,paperheight=27.0cm,top=2.0cm,bottom=2.0cm,left=1.5cm,right=1.5cm]{geometry}
\usepackage[switch]{lineno}
\usepackage{fancyhdr}
\begin{document}
%\begin{CJK*}{UTF8}{song}

\title{The new concepts of ether and calculation of the cosmological constant }
\author{Xiao-Song Wang}
\setcounter{footnote}{0}
\affil{{\normalsize Institute of Mechanical and Power Engineering, Henan Polytechnic University, Jiaozuo, Henan Province, 454000, China}}%
%\date{}
\date{Apr. 14th, 2024}
%\maketitle

%%%%%%%%%%%%% line numbering
%\linenumbers
%%%%%%%%%%%%% line numbering
\twocolumn [
	\begin{@twocolumnfalse}
		\maketitle
		\begin{abstract}
			\newgeometry{left=1.0cm, right=1.0cm}%
Since the general theory of relativity (GR) meets some difficulties, it seems that new considerations on the ether theories of gravitation in the history are needed. A theory of gravity based on some new concepts of ether and particles is briefly reviewed. In this theory, the universe is filled with a kind of fluid which may be called the $\Omega(0)$ substratum, or we say the gravitational ether. Particles are modeled as sink flows in the $\Omega(0)$ substratum. Newton's law of gravitation is derived by methods of fluid mechanics. Thus, gravity is interpreted as attractive force between sinks in the $\Omega(0)$ substratum. The theoretical calculation of the cosmological constant (CC) based on a mechanical model of vacuum is briefly reviewed. A proposed solution of the cosmological constant problem (CCP) is discussed. Inspired by the association of the gravitational wave (GW) event GW170817 and the gamma-ray burst (GRB) event GRB 170817A, we propose a theoretical calculation of the mass density of the electromagnetic ether.

\

keywords: cosmological constant; general theory of relativity; cosmological constant problem; vacuum mechanics; electromagnetic ether; gravitational wave.
	  \end{abstract}
	\end{@twocolumnfalse}
%abstract
%\vspace*{6pt}

\

%\noindent

%\thanks{$*$ E-mail: }

\

\

]

%%%%%%%%%%%%%%%%%%%%%%%%%%%%%%%%%%%%%%%%%%%%%%%%%%%%%%%%%%%%%%%%%%%%%%%%%%
\section*{Introduction \label{sec 100}}

The Einstein's field equations of gravity is a fundamental assumption in GR \cite{MisnerC1973}. Although GR has held up under every experimental test, it still face some difficulties \cite{WangXS201908,WangXS202104,WangXS202208}, for instance, medium of gravity, inharmonious between GR and quantum mechanics, CCP, the paradoxes of black holes, the velocity of the propagation of gravity, the definition of inertial system, origin of inertial force, gravitational waves, the speed of light in vacuum, the velocity of individual photons, etc. New considerations on the old concept of gravitational ether in the history may be needed.

If the cosmological term is absent in Einstein's equations, then a non-permanent universe is possible. However, this non-permanent picture of the universe contradicts with the philosophical belief that the universe endure from everlasting to everlasting (\cite{MisnerC1973}, p.\ 410). Therefore, in 1917 A. Einstein thought that the cosmological term should be added in his equations (\cite{MisnerC1973}, p.\ 410). In 1930, Hubble discovered the expansion of the universe (\cite{MisnerC1973}, p.\ 410). Thus, the cosmological term seems unnecessary. Einstein calling CC as the biggest blunder of his life. Thus, he abandoned the cosmological term and returned to his original equations (\cite{MisnerC1973}, p.\ 410). Later, CC was continuously and intensively studied \cite{ORaifeartaighC2018}.

Recently, we show that CC can be calculated theoretically based on a mechanical model of vacuum \cite{WangXS202208}. The predicted value of CC is in agreement with the observational value \cite{WangXS202208}. In this paper, we briefly review the theoretical calculation of CC in Ref. \cite{WangXS202208} and discuss CCP. The observation of the possible GW150914/GBM transient 150914 association \cite{LiX2016} and the association of GW170817 and GRB 170817A \cite{AbbottBP2017} suggest that  the hypothetical $\Omega(2)$ substratum in the previous model of vacuum \cite{WangXS202208} seems to be unnecessary. Thus, the mass density of the electromagnetic ether is obtained theoretically in this paper.

\section*{Brief review of some ether  theories of gravity in the history \label{sec 200}}
According to E. Whittaker, Descartes was the first to bring the concept of ether into science by suggesting that it has mechanical properties (\cite{WhittakerE1910}, p.\ 2).  Descartes interpreted the celestial motions of celestial bodies based on the hypothesis that the universe is filled by a fluidic vortex ether. He thought that the sun is the centre of an immense vortex formed of the first or subtlest kind of matter (\cite{WhittakerE1910}, p.\ 5). The vehicle of light in interplanetary space is matter of the second kind. Pressure is transmitted from a luminous object to the eye by the second kind of matter. Light is the transmission of this pressure.

Since Newton's law of gravitation was published in 1687, this action-at-a-distance theory was criticized by the French Cartesian. Newton pointed out that his inverse-square law of gravitation did not touch on the mechanism of gravitation (\cite{WhittakerE1951}, p.\ 28; \cite{HirosigeT1968}, p.\ 91). He tried to obtain a derivation of his law based on Descartes' scientific research program. At last, he proved that Descartes' vortex ether hypothesis could not explain celestial motions properly. Newton suggested an explanation of gravity based on the action of an etherial medium pervading the space (\cite{WhittakerE1951}, p.\ 28).

In the years 1905-1916, Einstein abandoned the concepts of electromagnetic ether and gravitational ether in his theory of relativity (\cite{Einstein1905}; \cite{Kostro2000}, p.\ 27-61). However, Einstein's assertion did not cease the explorations of ether. H. A. Lorentz believed that GR could be reconciled with the concept of an ether at rest and wrote a letter to A. Einstein (\cite{Kostro2000}, p.\ 65). Einstein changed his view later and introduced his new concept of ether (\cite{Kostro2000}, p.\ 63-113).

In 1920, Einstein said (\cite{Kostro2000}, p.\ 98):"{\itshape According to the general theory of relativity, space is endowed with physical qualities; in this sense, therefore, there exists an ether. According to the general theory of relativity, space without ether is unthinkable;}".

A. Einstein and L. Infeld said (\cite{EinsteinAInfeldL}, p.\ 256-257):"{\itshape Matter is where the concentration of energy is great, field where the concentration of energy is small. $\cdots$ What impresses our senses as matter is really a great concentration of energy into a comparatively small space. We could regard matter as the regions in space where the field is extremely strong.}"

In 1954, Einstein said (\cite{Kostro2000}, p149):"{\itshape There is no such thing as an empty space, i.e., a space without field. Space-time does not claim existence on its own, but only as a structural quality of the field.}"

\section*{Gravity is interpreted as attractive force between sinks by vacuum mechanics (VM)  \label{sec 500}}
In order to compare fluid motions with electric fields, J. C. Maxwell introduced an analogy between source or sink flows and electric charges (\cite{WhittakerE1951}, p.\ 243).

B. Riemann speculates that (\cite{RiemannB2004}, p.\ 507):"{\itshape  I make the hypothesis that space is filled with a substance which continually flows into ponderable atoms, and vanishes there from the world of phenomena, the corporeal world}".

H. Poincar$\acute{e}$ also suggests that matters may be holes in fluidic ether (\cite{PoincareH1997}, p.\ 171).

John C. Taylor proposed an idea that the inverse-square law of gravitation may be explained based on the concept of source or sink (\cite{TaylorJ2001}, p.\ 432).

Inspired by these sink flow models in the history, we suppose that the universe is filled by an ideal fluid which may be called the $\Omega(0)$ substratum \cite{WangXS200810}. We propose that microscopic particles are sink flows in the $\Omega(0)$ substratum \cite{WangXS200810}. Molecular are constructed by atoms. Atoms are formed by elementary particles. All the microscopic particles were made up of a kind of elementary sinks of the $\Omega(0)$ substratum \cite{WangXS200810}. These elementary sinks of the $\Omega(0)$ substratum may be called monads after Leibniz. These monads were created simultaneously. The initial masses and the strengths of the monads are the same. There exists the following attractive fore between two point sinks in the $\Omega(0)$ substratum \cite{WangXS200810}
\begin{equation}\label{attractive 300-100}
\mathbf{F}_{12}=-\rho_{0}\frac{Q_{1}Q_{2}}{4\pi r^{2}}\,\hat{\mathbf{r}}_{12},
\end{equation}
where $Q_{1}$ and $Q_{2}$ are the strengths of two sinks, $\mathbf{F}_{12}$ is the force exerted on the sink with strength $Q_{2}$ by another sink with strength $Q_{1}$, $\rho_{0}$ is the mass density of the $\Omega(0)$ substratum, $\hat{\mathbf{r}}_{12}$ denotes the unit vector directed along the line from the sink with strength $Q_{1}$ to the sink with strength $Q_{2}$, $r$ is the distance between the two sinks.

Using Eq.\  (\ref{attractive 300-100}), we show that the force $\mathbf{F}_{12}(t)$  exerted on the particle with mass $m_{2}(t)$ by the velocity field of the $\Omega(0)$ substratum induced by the particle with inertial mass $m_{1}(t)$ is \cite{WangXS200810}
\begin{equation}\label{gravitation 820-200}
\mathbf{F}_{12}(t)=- \gamma_{N}(t)\frac{m_{1}(t)
m_{2}(t)}{r^{2}}\hat{\mathbf{r}}_{12},
\end{equation}
where
\begin{equation}\label{constant 820-210}
\gamma_{N}(t)=\frac{\rho_{0} q^{2}_{0}}{4\pi m^{2}_{0}(t)},
\end{equation}
$m_{0}(t)$ is the inertial mass of a monad at time $t$, $-q_{0}( q_{0} > 0)$ is the strength of a monad.

Eq.\  (\ref{gravitation 820-200}) is similar to Newton's inverse-square-law of gravitation. We suppose that the parameter $\gamma_{N}(t)$ and the masses of particles are changing so slowly relative to the time scale of human beings that they can be treated as constants approximately. Thus, Newton's inverse-square law of gravitation may be regarded as a corollary of Eq.\  (\ref{gravitation 820-200}). Therefore, gravitation is interpreted as attractive force between sinks in the $\Omega(0)$ substratum \cite{WangXS200810}.

Recently, we speculate that gravitational phenomena in Fock's harmonic reference frames may be similar to those in inertial reference frames \cite{WangXS202104}. Following this research route, generalized Einstein's equations in some special non-inertial reference frames are derived \cite{WangXS202104}. If the field is weak and the reference frame is quasi-inertial, these generalized Einstein's equations reduce to Einstein's equations \cite{WangXS202104}. Thus, all the experiments which support GR may also support this theory.  For convenience, we may call this theory \cite{WangXS200810,WangXS201908,WangXS202104,WangXS202208} as the theory of vacuum mechanics (VM). A brief introduction of VM can be found in the appendix of Ref. \cite{WangXS202208}.

\section*{Review of the calculation of CC based on VM \label{sec 900}}
In 1990-1999 two groups found that some high redshift supernovae appeared fainter and thus more distant than they should be in a gravitationally decelerating universe (\cite{ByrdG2012}, p.\ 113). This discovery provides the first clue that the expansion of the universe is accelerating.

The concept of dark energy commonly denotes a catch-all term for the origin of the observed acceleration of the universe (\cite{ZylaPA2020}, p.\ 490). The first possible candidate of dark energy is that vacuum may contain some kind of substratum which behaves like Einstein's cosmological constant $\Lambda$ (\cite{ByrdG2012}, p.\ 113). The second possible explanation of dark energy is a modification of GR. The third possibility is that there may exist other unknown reasons to explain dark energy.

We focus on the first possibility, i.e., CC may stem from some substrata in vacuum. Lord Kelvin believes that the electromagnetic ether must also generate gravity \cite{KelvinL1901}. Presently we have no observational data of the density of the electromagnetic ether, or we call the $\Omega(1)$ substratum \cite{WangXS200804}. Therefore, there may exist the following two research routes. The first route is that the mass density $\rho_{1}$ of the $\Omega(1)$ substratum is exactly equal to the mass density $\rho_{\Lambda}$ corresponding to CC. The second route is that, except the $\Omega(0)$ and $\Omega(1)$ substratum, there exists a third kind of substratum in vacuum.

Presently, we cannot exclude or conclude the existence of a third kind of continuously distributed medium in the universe. Therefore, in Ref. \cite{WangXS202208} we tentatively introduce hypothetical $\Omega(2)$ substratum.

Recently, researchers noticed that the possible GW150914/GBM transient 150914 association and the association of GW170817 and GRB 170817A may have shed new light on fundamental physics \cite{LiX2016,AbbottBP2017}. In 2016, X. Li and et al. propose that if the possible GW150914/GBM transient 150914 association was confirmed, then this observation would provide the first opportunity to directly measure the velocity of GW \cite{LiX2016}. Further, the estimated difference between the velocity of GW and the speed of the light in vacuum should be within a factor of $\sim 10^{-17}$ \cite{LiX2016}. On August 17th 2017, GW event GW170817 was observed by the Advanced LIGO and Virgo detectors \cite{AbbottBP2017}. The GRB event GRB 170817A was observed independently by the Fermi Gamma-ray Burst Monitor and the Anti-Coincidence Shield for the Spectrometer for the International Gamma-Ray Astrophysics Laboratory \cite{AbbottBP2017}. The observed time delay of $+1.74 \pm 0.05 s$ between GRB 170817A and GW170817 shows that the difference between the speed $c_{gw}$ of GW and the speed of light is limited between $- 3 \times 10^{-15}c$ and $+ 7 \times 10^{-16}c$, where $c$ is the speed of light in vacuum \cite{AbbottBP2017}.

According to VM \cite{WangXS202104}, GW is the propagations of tensorial potential of gravitational fields in vacuum. If the speed $c_{gw}$ of GW is the same as the speed of light in vacuum, then $c_{gw}$ coincides with the speed of transverse elastic wave in the $\Omega(1)$ substratum. Thus, the $\Omega(1)$ substratum, or we say the electromagnetic ether, is the medium which propagates the tensorial potential of gravitational fields. Therefore, the hypothetical $\Omega(2)$ substratum in Ref. \cite{WangXS202208} seems to be unnecessary.

If the reference frame is quasi-inertial and the gravitational field is weak, then the generalized Einstein's equations in VM reduce to \cite{WangXS202104}
\begin{equation}\label{Einstein 150-100}
R_{\mu\nu}-\frac{1}{2}g_{\mu\nu}R = \frac{\kappa_{0}}{g_{0}}\left ( T^{\mathrm{m}}_{\mu\nu}+T^{\Omega(1)}_{\mu\nu}\right ),
\end{equation}
where $g_{\mu\nu}$ is the metric tensor of a Riemannian spacetime,
$R_{\mu\nu}$ is the Ricci tensor, $R\equiv g^{\mu\nu}R_{\mu\nu}$ is the scalar curvature, $g^{\mu\nu}$ is the contravariant metric tensor, $\kappa_{0}$ is Einstein's gravitational constant, $T^{\mathrm{m}}_{\mu\nu}$, $T^{\Omega(1)}_{\mu\nu}$ are the energy-momentum tensors of the matter system and the $\Omega(1)$ substratum respectively, $g_{0} \equiv \mathrm{Det} \ g_{\mu\nu}$.

The parameters $\kappa_{0}$ and $g_{0}$ depend on the choice of coordinate system. Following V. Fock (\cite{FockV1964}, p.\ 195), we choose the coordinate system $S_{0} \equiv \{t, x, y, z\}$. Since there are no atoms in vacuum, the generalized Einstein's equations Eqs.\ (\ref{Einstein 150-100}) in vacuum reduce to
\begin{equation}\label{Einstein 800-1400}
R^{\mu\nu}-\frac{1}{2}g^{\mu\nu}R = \frac{\kappa_{0}}{g_{0}} T_{\Omega(1)}^{\mu\nu},
\end{equation}
where $R^{\mu\nu}$ is the contravariant Ricci tensor, $T_{\Omega(1)}^{\mu\nu}$ is the contravariant energy-momentum tensors of the $\Omega(1)$ substratum.

In the coordinate system $S_{0}$, the Einstein's gravitational constant $\kappa_{0}$ can be written as \cite{WangXS202208}
\begin{equation}\label{kappa 800-2000}
\kappa_{0}=\frac{8\pi G}{c^{2}},
\end{equation}
where $G$ is Newton's gravitational constant, $c$ is the velocity of light in vacuum.

We speculate that the cosmological term may stem from the term on the right hand side of Eqs.\ (\ref{Einstein 800-1400}) \cite{WangXS202208}. Applying a theorem of V. Fock on the mass tensor of a fluid, we obtain the contravariant energy-momentum tensor $T_{\Omega(1)}^{\mu\nu}$ of the $\Omega(1)$ substratum \cite{WangXS202208}. Solving the the field equations (\ref{Einstein 800-1400}), we get the approximate value of the contravariant metric tensor $g^{\mu\nu}$ \cite{WangXS202208}. Introducing some auxiliary assumptions, we obtain the following relations \cite{WangXS202208}
\begin{equation}\label{relationship 1100-400}
\frac{\kappa_{0}}{g_{0}} T_{\Omega(1)}^{\mu\nu} \approx -\kappa_{0}\rho_{1}g^{\mu\nu},
\end{equation}
where $\rho_{1}$ is the rest mass densities of the $\Omega(1)$ substratum in a laboratory frame.

We introduce the following notation
\begin{equation}\label{notation 1100-100}
\Lambda = \kappa_{0}\rho_{1}.
\end{equation}

Using Eq.\ (\ref{notation 1100-100}), Eqs.\  (\ref{relationship 1100-400}) can be written as
\begin{equation}\label{relationship 1100-500}
\frac{\kappa_{0}}{g_{0}}T^{\Omega(1)}_{\mu\nu} \approx -\Lambda g_{\mu\nu}.
\end{equation}

We notice that the term $-\Lambda g_{\mu\nu}$ in Eqs.\  (\ref{relationship 1100-500}) coincides with the cosmological term in Einstein's field equations (\cite{MisnerC1973}, p.\ 410). Using Eqs.\  (\ref{relationship 1100-500}), Eqs.\  (\ref{Einstein 150-100}) can be written as
\begin{equation}\label{Einstein 900-200}
R_{\mu\nu}-\frac{1}{2}g_{\mu\nu}R = \frac{\kappa_{0}}{g_{0}} T^{\mathrm{m}}_{\mu\nu}-\Lambda g_{\mu\nu}.
\end{equation}

Eqs.\  (\ref{Einstein 900-200}) are generalized equations of Einstein's field equations with the cosmological term.

Comparing Eq.\  (\ref{notation 1100-100}) and Eq.\ (\ref{kappa 800-2000}), we have
\begin{equation}\label{relationship 1100-700}
\Lambda = \frac{8\pi G \rho_{1}}{c^{2}}.
\end{equation}

The theoretical value of CC $\Lambda_{\mathrm{the}}$ is \cite{WangXS202208}
\begin{equation}\label{Lambda 1100-1200}
\Lambda_{\mathrm{the}} = 1.093(65)\times 10^{-52}\mathrm{m}^{-2}.
\end{equation}

The theoretical value of CC $\Lambda_{\mathrm{the}}$ in Eq.\  (\ref{Lambda 1100-1200}) is consistent with the observational value of CC $\Lambda_{\mathrm{obs}}=1.088(30)\times 10^{-52}\mathrm{m}^{-2}$ (\cite{ZylaPA2020}, p.\ 138). A comparison of the theoretical and the observational values of CC can be found in Table \ref{cosmological 900-100}.
\begin{table}
\caption{Comparison of the theoretical and the observational values of CC. $\Lambda_{\mathrm{the}}$ is the theoretical value of CC. $\Lambda_{\mathrm{obs}}$ is the observational value of CC.}
\begin{tabular}{|c|c|c|}
\hline
 & data & reference \\
\hline
$\Lambda_{\mathrm{the}}$ & $1.093(65)\cdot 10^{-52}(\mathrm{m}^{-2})$ & \cite{WangXS202208} \cr
\hline
$\Lambda_{\mathrm{obs}}$ & $1.088(30)\cdot 10^{-52}(\mathrm{m}^{-2})$ & \cite{ZylaPA2020}, p.\ 138  \cr
\hline
\end{tabular}
\label{cosmological 900-100}
%\footnotetext{}
\end{table}

We have shown that the origin of the cosmological term $\Lambda g_{\mu\nu}$ in Eqs.\ (\ref{Einstein 900-200}) is the energy-momentum tensors $T^{\Omega(1)}_{\mu\nu}$ of the $\Omega(1)$ substratum \cite{WangXS202208}. Therefore, we speculate that the $\Omega(1)$ substratum may be a possible candidate of the so-called concept of dark energy (\cite{ZylaPA2020}, p.\ 490).

\section*{A possible solution of CCP \label{sec 1000}}
In 1968, Y. B. Zeldovich suggested a lower bound $\Lambda_{QFT}=10^{-6}\mathrm{m}^{-2}$ of CC, corresponding to a mass density of $\rho_{\Lambda}=10^{20} \mathrm{kg}\cdot\mathrm{m}^{-3}$ \cite{ZeldovichYB1968,SahniV2008,ORaifeartaighC2018}. However, an observational data of CC is $\Lambda_{obs}=1.088(30)\times 10^{-52}\mathrm{m}^{-2}$ (\cite{ZylaPA2020}, p.\ 138), corresponding to a mass density of $\rho_{\Lambda}=5.831(02)\times 10^{-27} \mathrm{kg}\cdot\mathrm{m}^{-3}$. Thus, quantum field theory (QFT) predicted a value $\Lambda_{QFT}$ of CC that was $46$ orders of magnitude larger than that observed. This theoretical problem is known as CCP \cite{ORaifeartaighC2018}.

The first class of solutions of CCP is to modify the theory of gravitation. The second class of solutions is to revise the standard model of particle physics. However, the CCP are still open \cite{ORaifeartaighC2018}.

The origin of CCP may be that GR is a phenomenological theory of gravity \cite{WangXS202208}. From the viewpoint of VM \cite{WangXS201908,WangXS202104}, only those energy-momentum tensor of sink flows in the $\Omega(0)$ substratum, i.e. the energy-momentum tensor $T^{\mathrm{m}}_{\mu\nu}$ of matter and the energy-momentum tensor $T^{\Omega(1)}_{\mu\nu}$ of the $\Omega(1)$ substratum are qualified for the source terms in the generalized Einstein's equations. Not all kinds of energy-momentum tensors are permitted to act as source terms in the generalized Einstein's equations. Therefore, the zero-point energy of electromagnetic fields, the energy from the electro-weak phase transition, the energy from the quantum chromodynamic phase transition, etc., should not act as source terms in the generalized Einstein's equations. Thus, the cosmological term $-\Lambda g_{\mu\nu}$ does not result from the zero-point energy of electromagnetic fields or other energies.

\section*{Calculation of the mass density of the electromagnetic ether \label{sec 1100}}
The mass density of the electromagnetic ether remains unknown since eighteenth century \cite{WhittakerE1953
}. Since the hypothetical $\Omega(2)$ substratum \cite{WangXS202208} may be unnecessary, we set $\rho_{2} =0$ in Eq.\  (107) in Ref. \cite{WangXS202208} and obtain
\begin{equation}\label{density 2000-100}
\rho_{1}  = 5.831(02)\times 10^{-27} \mathrm{kg}\cdot\mathrm{m}^{-3}.
\end{equation}

Eq.\  (\ref{density 2000-100}) is the theoretically predicted value of the mass density of the electromagnetic ether based on VM. It is interesting whether it is possible for us to carry out experiments or observations to test this prediction.

\section*{Differences between VM and GR \label{sec 1200}}
There exist some differences between VM and GR \cite{WangXS202104}, refers to Table \ref{difference 800-100}.
\begin{table}
\caption{Differences between VM and GR}
\begin{tabular}{|l|l|l|l|}
\hline
& GR & VM \\
\hline
{\tiny field equations} & {\tiny Einstein's equations (EE)} & {\tiny generalized EE} \cr
\hline
{\tiny field equations} & {\tiny assumptions} & {\tiny mechanics} \cr
\hline
{\tiny reference frames} & {\tiny all reference frames} & {\tiny Fock system} \cr
\hline
{\tiny Einstein's equations} & {\tiny rigorous} & {\tiny approximation} \cr
\hline
{\tiny medium of gravity} &  {\tiny no medium} & {\tiny $\Omega (0)$ substratum} \cr
\hline
{\tiny theory type} & {\tiny phenomenological} & {\tiny mechanics} \cr
\hline
{\tiny Riemannian spacetime} & {\tiny an assumption} & {\tiny defined} \cr
\hline
{\tiny metric tensor} & {\tiny an assumption} & {\tiny defined} \cr
\hline
{\tiny masses of particles} & {\tiny constants} & {\tiny variable} \cr
\hline
{\tiny gravitational constant} & {\tiny constant} & {\tiny variable} \cr
\hline
{\tiny adjustable parameters} & {\tiny no} & {\tiny yes} \cr
\hline
\end{tabular}
\label{difference 800-100}
%\footnotetext{}
\end{table}

The Einstein's equations are supposed to be valid in all reference frames \cite{MisnerC1973}. In VM, the generalized Einstein's equations are only valid in the Fock coordinate systems \cite{WangXS202104}.  Experimental tests of GR are carried out only in the solar system \cite{WillCW2014}. The solar system can be approximately regarded as a quasi-inertial reference frame \cite{WangXS202104}. Therefore, it is still not clear whether the Einstein's equations are valid in all non-inertial reference frames or not.

It may be valuable for us to carry out possible experiments or observations to detect some of these differences between VM and GR.

\section*{Conclusion \label{sec 3000}}
Some ether  theories of gravity in the history is briefly reviewed. Then, we discuss a recently proposed theory of gravitation based on some new concepts of ether and particles. In this theory, the universe is filled with a kind of fluid which may be called the $\Omega(0)$ substratum, or we say the gravitational ether. Particles are modeled as sink flows in the $\Omega(0)$ substratum. Thus, Newton's inverse-square law of gravitation is derived by methods of hydrodynamics based on the sink flow model of particles. Generalized Einstein's equations in the Fock coordinate systems are derived. Following Lord Kelvin, we suppose that the electromagnetic ether, or we call the $\Omega(1)$ substratum, may also generate gravity. Thus, CC is calculated theoretically. The predicted value of CC is consistent with the observational value. The $\Omega(1)$ substratum may be a possible candidate of the dark energy. According to VM, only those energy-momentum tensors of sinks in the $\Omega(0)$ substratum are permitted to act as the source terms in the generalized Einstein's equations. Other kinds of energy-momentum tensors are not allowed to act as source terms in the generalized Einstein's equations. This is a proposed solution of CCP based on VM. The observed time delay of $+1.74 \pm 0.05 s$ between GRB 170817A and GW170817 shows that the speed of GW equals the speed of light in vacuum. Therefore, the hypothetical $\Omega(2)$ substratum seems to be unnecessary. Thus, the mass density of the electromagnetic ether is calculated theoretically.

%\nocite{*}

%\end{CJK*}
\end{document}